# Performance of three model-based iterative reconstruction algorithms using a CT task-based image quality metric.

Gaia Muti[1,2], Stefano Riga[2], Luca Berta[2], Denise Curto[1,2], Cristina De Mattia[2], Marco Felisi[2], Francesco Rizzetto[3], Alberto Torresin[2], Angelo Vanzulli[3,4], Paola Enrica Colombo[2]

[1] Department of Physics, Università degli Studi di Milano, Milan, Italy
[2] Medical Physics Department, ASST Grande Ospedale Metropolitano Niguarda, Milano, Italy
[3] Department of Radiology, ASST Grande Ospedale Metropolitano Niguarda, Milan, Italy
[4] Department of Oncology and Hemato-Oncology, Università degli Studi di Milano, Milan, Italy

**Abstract**

**Objectives** To evaluate and compare the task-based image quality of a low contrast clinical task for the abdomen protocol (e.g., pancreatic tumour) of three different CT vendors, exploiting three model-based iterative reconstruction (MBIR) levels.

**Methods** We used three CT systems equipped with a full, partial, advanced MBIR algorithms. Acquisitions were performed on a phantom at three dose levels. Acquisitions were reconstructed with a standard kernel, using filtered back projection algorithm (FBP) and three levels of the MBIR. The noise power spectrum (NPS), the normalized one (nNPS) and the task-based transfer function (TTF) were computed following the method proposed by the American Association of Physicists in Medicine task group report-233 (AAPM TG-233). Detectability index (d') of a small lesion (small feature; 100 HU and 5-mm diameter) was calculated using non-prewhitening with eye-filter model observer (NPWE).

**Results** The nNPS, NPS and TTF changed differently depending on CT system. Higher values of d' were obtained with advanced-MBIR, followed by full-MBIR and partial-MBIR.

**Conclusion** Task-based image quality was assessed for three CT scanners of different vendors, considering a clinical question. Detectability can be a tool for protocol optimisation and dose reduction since the same dose levels on different scanners correspond to different d' values.

**Abbreviations**
CTDIvol Volume CT dose index, ESF Edge-spread function, FBP Filtered back projection, LSF Line-spread function, MBIR model-based iterative reconstruction, nNPS Normalized noise power spectrum, NPS Noise power spectrum, NPWE Non-prewhitening with eye-filter, TTF Task-based transfer function

**Introduction**

The use of computed tomography (CT) holds a key role in the diagnosis and follow-up of several pathologies. There are several factors which make CT images preferable in many clinical scenarios, such as wide availability, speed of response, and good performance; however, scan involves exposure to ionizing radiation. For this reason, both vendors, medical physicist expert and clinicians are continuously working to optimise acquisition protocols. The main goal of optimization process is obtaining the best image quality by reducing the dose to acceptable values for different clinical tasks. On vendor side, this effort led to the introduction of new different technologies, such as image reconstruction.

In the early 2010s, CT scanners with statistical iterative reconstruction algorithms (IR, first-generation) were introduced; they were able to improve image quality allowing a first reduction of patient dose [1]. Over the next six years, model-based iterative reconstruction algorithms (MBIR, second-generation) were then presented [2]; these algorithms use a probabilistic method to reduce noise and artifacts [1]. MBIR can be divided in partial, full and advanced. Partial-MBIR is a hybrid between IR and MBIR technologies [3], full-MBIR uses knowledge-based iterative reconstruction constrained by a cost function [4] and advanced-MBIR uses an advanced regularization loop operating in a 3D-voxel neighbourhood to separates noise from actual anatomical structures to preserve natural anatomical texture appearance [5].

At the date of this study, the state-of-the-art is in third-generation CT with methods that leverage deep learning algorithm in the image reconstruction (DLIR), developed principally from two vendors, GE Healthcare iterative TrueFidelity and Canon Medical AiCE [6], not yet of common use in clinical practice [7,8,9].

On user side, with ICRP-73 (1996) [10] and later ICRP-135 (2017) [11], diagnostic reference levels (DRLs) were introduced as an optimisation tool to lower variability in clinical practice, pointing out the need to act if local dose level indicators exceed the DRLs. In 2021, with a Europe-wide survey, DRLs were collected by matching dose to a specific clinical indication (clinical task) (RP-195, EUCLID project) [12]. The limitation of this study was that it did not consider image quality, a key aspect of objectively rating CT examination protocol. In the last two decades, however, there were several studies to define which metrics should be used to evaluate image quality while considering clinical task. Starting from objective analysis of noise power spectrum [13-19] and spatial resolution [17,21,22,23], the model observer methodologies try to combine system image performance (amount of information), task characteristics and the degree of clinician image perception [24]. The model that has a better agreement with the human observer is obtained with the non-prewhitening matched filter (NPW) by adding the eye filter correction [25]. In this study, the detectability index (d'), that quantifies this degree of separation for signal present/signal absent distributions calculated as suggested by the American Association of Physicists in Medicine task group report-233 (AAPM TG-233) [26].

The d' metric is strongly related to the clinical task investigated. For example, Greffier reported d' values for the detection of pulmonary arteriovenous malformation [27], lytic and sclerotic bone lesions [28] and focal liver lesions [29].

This study evaluated and compared the task-based image quality of a low contrast clinical task for abdomen protocol of three different scanners for three MBIR levels and explored the potential dose reduction without quality loss. The aim is to assess, for scanners with different iterative algorithms within our hospital centre, a specific clinical protocol using a standardised metric.

**Materials and methods**

**CT systems, acquisition and reconstruction parameters** CT systems of three different vendors were studied: Revolution (GE Healthcare), Ingenuity (Philips Medical Systems), Somatom Drive (Siemens Healthineers), all equipped with MBIR. GE scanner installed a partial-MBIR (adaptive statistical iterative reconstruction-Veo, Asir-V), Philips installed a full-MBIR (iterative model reconstruction, IMR), and Siemens installed an advanced-MBIR (advanced modeled iterative reconstruction, ADMIRE) [2]. We used a common institutional clinical CT protocol dedicated to different abdominal diagnostic tasks with a standard dose level (volume CT dose indexes: $CTDI_{vol} \sim 13 mGy$), disabling tube current modulation for the phantom imaging. The same protocol was acquired with two other dose levels, reduced dose and low dose to explore potential dose reduction due to MBIR. The reduced and low dose was 60% and 30% of the standard value, respectively. $CTDI_{vol}$ determined for a 32-cm diameter reference phantom, was estimated according to recommendations from the International Electrotechnical Commission (IEC 60601-2-44) and retrieved from radiation-dose structured reports. Standard kernel was used in all systems. Acquisitions were reconstructed with the filtered back projection (FBP) and with three increasing strength levels for each MBIR, chosen to explore the whole range of possibilities offered by different CT models. We referred to them as low, medium and high levels. Settings for CT data acquisitions and image reconstructions were reported in Tab.1; the most similar settings were chosen for the different scanners. The total dataset of CT images consisted of 36 different cases available for analysis: a combination of three CT systems (Revolution, Ingenuity, Somatom Drive), three dose levels (standard, reduced and low) and four reconstruction methods (FBP, low, medium and high MBIR).

|  | **Revolution GE** | **Ingenuity Philips** | **DRIVE Siemens** |
|---|---|---|---|
| **Dose level** |  |  |  |
| Standard CTDIvol (mGy) | 13.26 | 13.10 | 13.42 |
| Reduced CTDIvol (mGy) | 8.84 | 8.80 | 8.94 |
| Low CTDIvol (mGy) | 4.42 | 4.30 | 4.47 |
| **Data acquisition** |  |  |  |
| Tube potential (kVp) | 120 | 120 | 120 |
| Gantry revolution time (s) | 0.5 | 0.5 | 0.5 |
| Beam collimation (mm) | 0.625x128 | 0.625x64 | 0.6x64 |
| Pitch | 0.992 | 1 | 1 |
| Field of View (cm) | 250 | 250 | 250 |
| **Image reconstruction** |  |  |  |
| Display field of view (cm) | 200 | 200 | 200 |
| Slice thickness (mm) | 2.5 | 2.5 | 3 |
| Kernel | STD | B | Br40 |
| MBIR | Asir-V | IMR | ADMIRE |
| MBIR level | 30%, 50%, 70% | 1, 2, 3 | 1, 3, 5 |

Tab.1 Acquisition (CT setting and dose level) and reconstruction parameters used for each CT scan

**Phantom** A Catphan® 600 (The Phantom Laboratory Inc., Salem, NY, USA) phantom was used (Fig.1-A) [30]. The CTP404 module (Fig.1-B) had eight sensitometry samples of 15-mm diameter and 25-mm thickness each placed into a uniform material. Sensitometry samples were Teflon, Delrin, Acrylic, Polystryene, LDPE, PMP and two of Air [30]. The Polystryene insert was used to evaluate the task-based transfer function (TTF). The module CTP486 (Fig.1-C), cast from a uniform material, with a CT number inside 2% (0-20 HU) of water, was used to compute the noise power spectrum (NPS).

**Image quality assessment** An open-source software (imQuest, Duke University, Durham, NC, USA) was used to evaluate the image quality. This user-friendly image analysis tool was developed to get task-based image quality of CT images as suggested by the AAPM TG-233 [26]. The main indexes calculated by imQuest are: TTF of selected insert, NPS for a selectable number of regions of interest (ROI) and d' through the definition of the task function and the model observer.

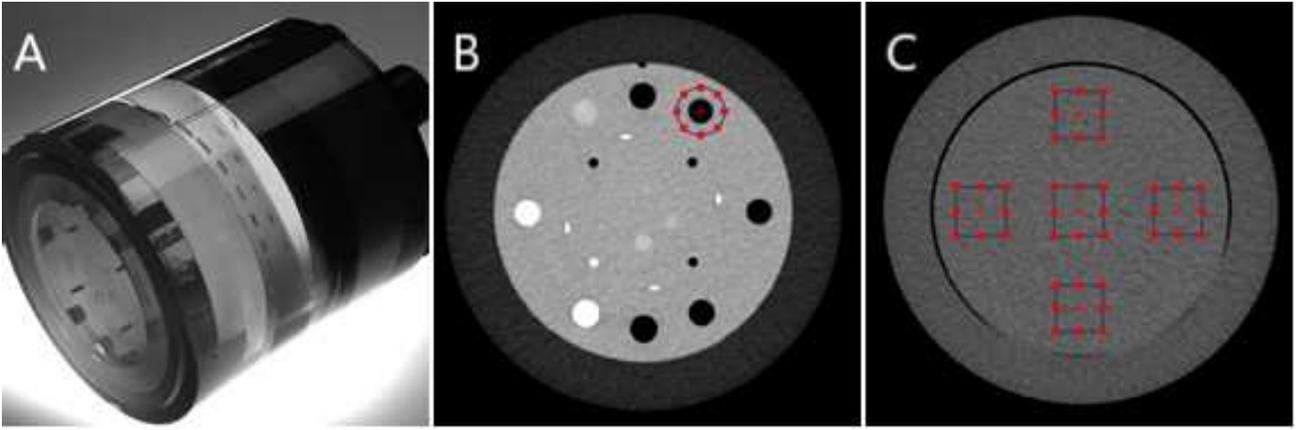

Fig. 1 A: CATPHAN 600, phantom. B: ROI used to compute the task-based transfer function (TTF). C: ROIs used for the noise power spectrum (NPS) calculation and normalized NPS calculation.

As well known, estimation of spatial resolution for iterative reconstruction, through Modulation Transfer Function (MTF), can suffer from limitations. A correct evaluation can be done using TTF measured from an object with the same specific contrast [21]. In CTP404 module, we chose the Polystyrene insert, which has an object-to-background contrast (|ΔHU|≈100) adequate for low-contrast diagnostic task. It corresponds to the one faced clinically by radiologists when evaluating abdominal and pelvic CT scans for suspected pancreatic tumour. We placed a circular ROI with a radius about twice the insert on the chosen insert (Fig.1-B); software measured radial edge-spread function (ESF) firstly and then ESF derivative, obtaining the line-spread function (LSF) and finally it evaluated TTF as the normalized Fourier transform of LSF. ESF was computed on 15 consecutives axial images to get a total effective contrast to noise ratio (CNR) greater than 15 to have a 10% error on TTF calculation [17]. Spatial resolution characteristics amongst different CT vendors, different reconstruction algorithms and MBIR levels were compared with FBP results, qualitatively in the shape of the TTF curve and quantitatively in terms of $TTF_{50\%}$ frequency shift respect to the FBP one. The shift was evaluated as the percentage difference of $TTF_{50\%}$ frequency value of the iterative algorithm compared to that of the FBP.

Noise was characterized by computing noise power spectrum (NPS) and the normalized one (nNPS). NPS was calculated drawing five 128×128 pixels ROIs within 15 consecutives axial images in the CTP486 module, Fig.1-C. Five ROIs were used to obtain a 10% error in NPS calculation [17]. The 2D-NPS was then computed as the area-normalized Fourier transform of ROI [31]. Then, exporting the NPS data from imQuest, nNPS was calculated normalizing the spectrum by its integral over the full range of spatial frequency. Noise characteristics were compared in terms of NPS area and nNPS shift peak frequency; peak close to low or high spatial frequencies denote a mottled or finer noise textures, respectively [32]. The shift was evaluated as the percentage difference in the peak frequency value of the iterative algorithm compared to that of the FBP. Since spatial resolution and noise both affect image quality, the d' metric allows to evaluate the jointly effect. To compare it, a specific clinical task was defined, synthesizing an ideal signal to be detected. We simulated a pancreatic lesion (5-mm diameter lesions with |ΔHU|≈100 HU with a Gaussian profile with blurring factor of 1) using the non-prewhitening with eye filter (NPWE) model observer with a radial noise generation mode and Saunders visual response function, which included filtering of the image with an eye-filter, intended to obtain a better agreement of the performance of this model and human observer [33]; d' was calculated using the following formula:

$$d'^2_{NPWE} = \frac{[\iint |W(u,v)|^2 TTF(u,v)^2 Z(u,v)^2 du\, dv\,]^2}{\iint |W(u,v)|^2 TTF(u,v)^2 NPS(u,v)^2 Z(u,v)^2 du\, dv}$$

where u and v were spatial frequencies in x and y directions, respectively, Z(u,v) was the eye filter that models the human visual system sensitivity to different spatial frequencies and W(u,v) was the task function defined as the Fourier transform of this synthesized signal [7,26]. For the eye filter imQuest required the setting to reproduce the working scenario of radiologists the following interpretation condition were considered: display pixel pitch of 0.2 mm, zoom factor of 1.74, viewing distance of 500-mm, and a field of view of 380 mm.

As a final step, the dose reduction that occurs using the iterative reconstruction algorithms was evaluated; this was calculated by extrapolating the dose value needed to achieve the same detectability as at the standard dose with FBP reconstruction.

**Results**

**Spatial Resolution** TTF trend for different levels of MBIR, for each dose level and scanner was shown in Fig.2. Frequency shifts of $TTF_{50\%}$ respect to the FBP values were reported in Tab.2, where a positive value denoted an increase from the FBP frequency. Graphs and table showed a scanner dependent trend of TTF as the MBIR strength increased. For GE scanner, TTF showed no change as the MBIR level increased. For Philips scanner, TTF curves for the iterative reconstructions deviated from the FBP one, showing a higher spatial resolution. This was reflected in $TTF_{50\%}$ shift: at standard dose, the shift from FBP was about 40%, while the curves at different levels of iterative were overlapping. For Siemens scanner there was a progressive rise in spatial resolution at each MBIR step, with a TTF$_{50\%}$ shift of more than 10% for the low iterative level, more than 20% for the medium and around 40% for the high level, independently of the dose value. As the dose varied, there was a limited shift in TTF$_{50\%}$ of less than ±10%; this value had the same order of magnitude to the metric error [17].

| | **GE** | | | **Philips** | | | **Siemens** | | |
|---|---|---|---|---|---|---|---|---|---|
| **Radiation dose level** | Standard | Reduced | Low | Standard | Reduced | Low | Standard | Reduced | Low |
| CTDI_vol (mGy) | 13.26 | 8.84 | 4.42 | 13.10 | 8.80 | 4.30 | 13.42 | 8.94 | 4.47 |
| **TTF$_{50\%}$** | | | | | | | | | |
| FBP *reference value*(mm$^{-1}$) | 0.39 | 0.41 | 0.42 | 0.33 | 0.33 | 0.32 | 0.37 | 0.39 | 0.41 |
| MBIR_low | 0% | 1% | -1% | 41% | 35% | 28% | 14% | 14% | 12% |
| MBIR_medium | 1% | 1% | -1% | 40% | 35% | 28% | 24% | 23% | 21% |
| MBIR_high | 1% | 1% | -2% | 41% | 35% | 28% | 43% | 38% | 37% |
| **NPS area** | | | | | | | | | |
| FBP *reference value*(HU) | 7.84 | 9.76 | 13.81 | 6.39 | 7.58 | 10.19 | 5.69 | 6.85 | 9.64 |
| MBIR_low | -22% | -22% | -22% | -32% | -35% | -35% | -10% | -10% | -10% |
| MBIR_medium | -36% | -36% | -36% | -47% | -50% | -52% | -29% | -30% | -30% |
| MBIR_high | -50% | -50% | -49% | -60% | -64% | -68% | -52% | -53% | -54% |
| **nNPS peak** | | | | | | | | | |
| FBP *reference value*(mm$^{-1}$) | 0.30 | 0.28 | 0.27 | 0.22 | 0.25 | 0.22 | 0.24 | 0.24 | 0.27 |
| MBIR_low | -26% | -28% | -18% | -57% | -44% | -36% | -7% | 0% | 0% |
| MBIR_medium | -42% | -33% | -29% | -57% | -63% | -57% | -7% | -7% | -22% |
| MBIR_high | -47% | -39% | -41% | -57% | -63% | -57% | -27% | -20% | -41% |
| **Detectability Index** | | | | | | | | | |
| FBP *reference value* | 29.32 | 23.14 | 16.17 | 31.07 | 27.21 | 18.21 | 45.38 | 36.32 | 27.12 |
| MBIR_low | 9% | 9% | 7% | 22% | 22% | 28% | 6% | 7% | 8% |
| MBIR_medium | 16% | 15% | 12% | 29% | 28% | 36% | 20% | 23% | 22% |
| MBIR_high | 22% | 21% | 18% | 36% | 36% | 46% | 42% | 48% | 43% |

Tab.2 Polystryene TTF50% frequency, noise area, NPS peak frequency and detectability index for FBP reconstruction, for each scanner, at three levels of radiation dose. For the iterative reconstruction, percentage differences in comparison with FBP were reported for three MBIR levels.

**Noise** The effect on NPS for different levels of MBIR, for each dose and scanner was shown in Tab.2. In Fig.3 each graph showed the nNPS of FBP and three increasing MBIR. Following the spectrum normalization, graphs of same MBIR at all dose values were similar and only minor differences were observed. Tab.2 reported the percentage differences in the NPS area and the frequency shift of the nNPS peak, moving from the FBP reconstruction to the MBIR ones, where a positive value denoted an increase compared with the FBP values.

At all three dose levels investigated and for each scanner, noise area markedly decreased increasing MBIR level, up to -68%. For FBP reconstruction, higher noise values were found for GE, followed by Philips and finally Siemens, which had the lowest noise area for each dose level examined. The three MBIRs reduced noise area differently. For Philips there was a greater percentage noise reduction from FBP for each iterative level, compared to the other scanners. Considering the highest MBIR level, Philips achieved a reduction of 60% at standard dose level, while GE and Siemens a reduction of 50%.

Reconstruction algorithm acted on the magnitude of noise, but also on the shape of the NPS curve. Philips scanner presented the greatest change of the nNPS with a shift of the peak till 63% towards the low spatial frequencies respect to the FBP value. The smallest changes in the spectrum shape occurred with Siemens scanner for standard and reduced dose level.

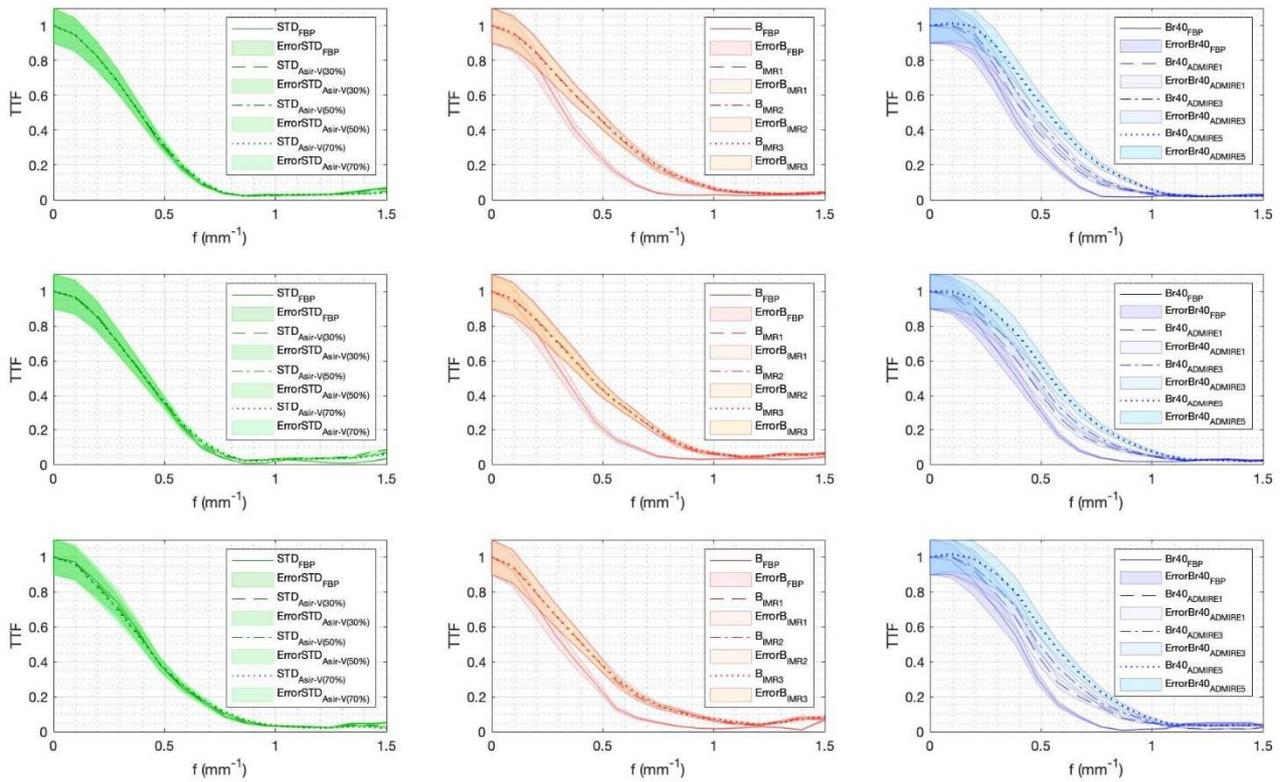

Fig. 2 TTF curves for polystryene insert with FBP and MBIR algorithms for three dose levels. Standard dose: first row, reduced dose: second row, low dose: third row. GE: green, Philips: orange, Siemens: blue

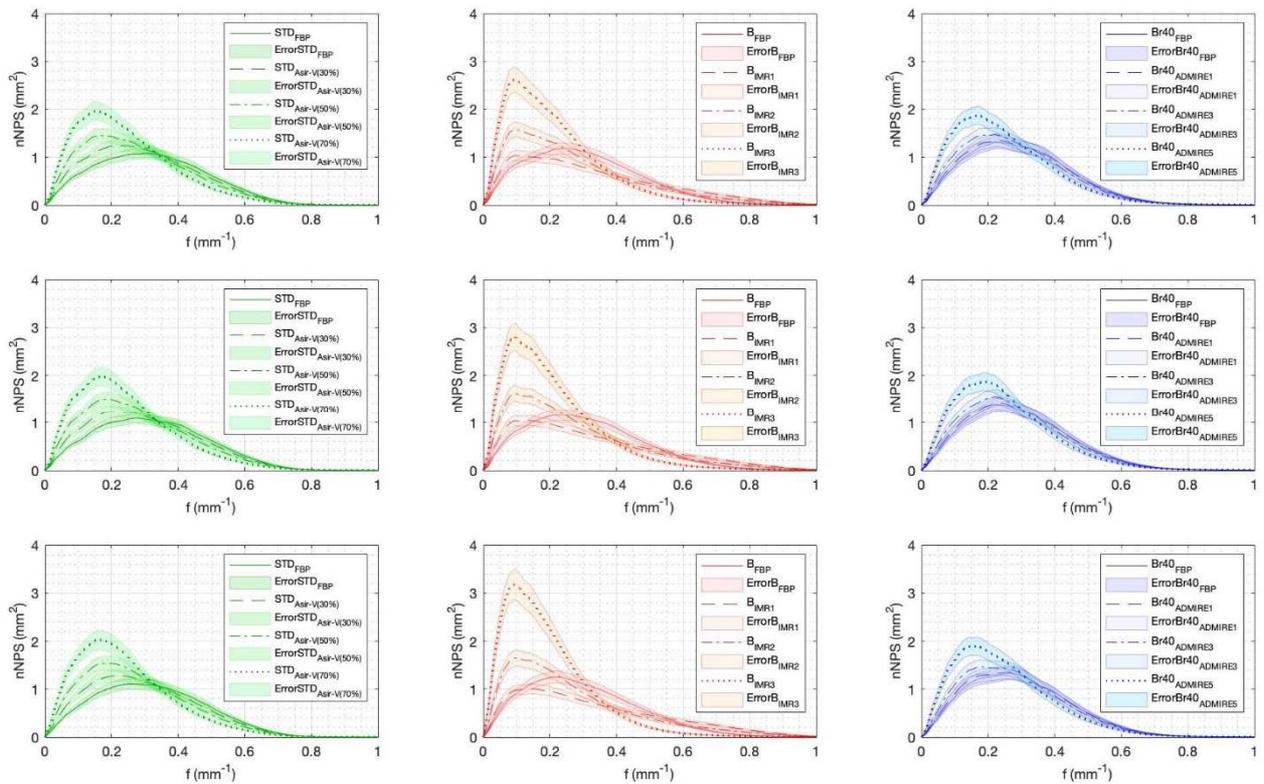

Fig. 3 nNPS curves with FBP and MBIR algorithms for three dose levels. Standard dose: first row, reduced dose: second row, low dose: third row. GE: green, Philips: orange, Siemens: blue

**Detectability index** At all three dose levels investigated and for each CT, the detectability increased with MBIR level. Tab.2 reported the percentage increments in d' relative to FBP values for each dose level and for each scanner. Siemens, which already started from a higher level of detectability in FBP, had the largest increase with the highest level of iterative, ranging from 42% to 48%, followed by Philips with an increase ranging from 36% to 46% and last GE with an increase of about 20%. Fig.4 showed the detectability values as a function of $CTDI_{vol}$ for each iterative level used for GE, Philips, and Siemens scanner. The figure showed that detectability increased with rising dose and iterative level for all scanners, but for Siemens this was more pronounced, achieving the highest values and in absolute terms.

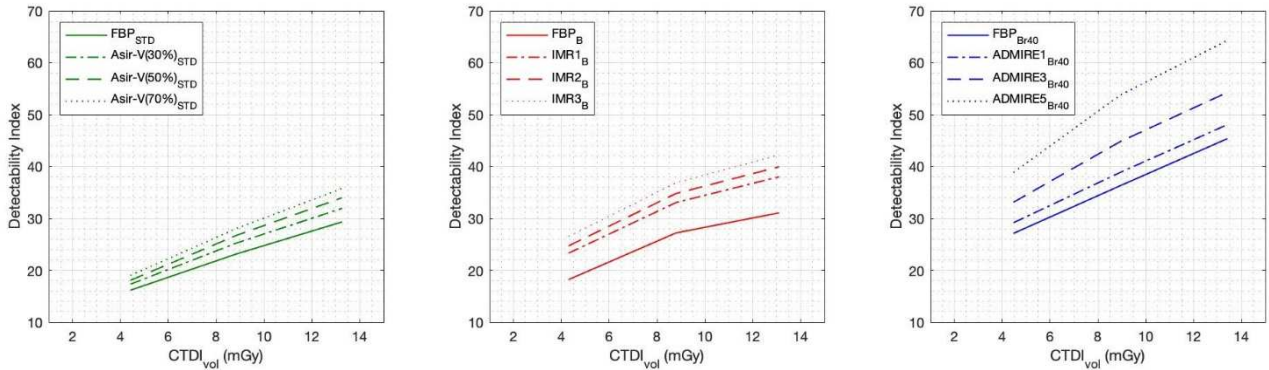

Fig. 4 Detectability index (d′) as a function of dose for detection of the small feature (5 mm in diameter, 100 HU contrast). GE: green, Philips: orange, Siemens: blue

**Dose reduction** To evaluate the dose saving, we extrapolated the $CTDI_{vol}$ needed to achieve the same d' value of FBP at standard dose using one of three investigated MBIR levels. Potential percentage dose reduction for each scanner was calculated in Tab.3.

|  | FBP detectability index at standard dose | Dose reduction (%) | | |
|---|---|---|---|---|
|  |  | **MBIR_low** | **MBIR_medium** | **MBIR_high** |
| **GE** | 29.32 | -13% | -21% | -28% |
| **Philips** | 31.07 | -40% | -46% | -54% |
| **Siemens** | 45.38 | -8% | -31% | -51% |

Tab.3 Percentage dose reduction with low/medium/high MBIR compared to FBP at standard dose.

Starting from a standard dose FBP protocol it was feasible to reduce the dose by the value reported in the Tab.3 by setting a low, medium, or high MBIR level. The table showed that the GE iterative algorithm allowed a limited dose reduction of -28% setting the highest MBIR level, Philips had the largest dose reduction gap when moving from FBP to MBIR. Siemens, which started from higher values in FBP, presented a limited reduction with the first level of MBIR (-8% vs. -40% for Philips) but showed a larger gap with the medium and high level of MBIR (-31% and -51%, respectively). The impact of MBIR and dose on d' were illustrated through colour-maps in Fig.5 to highlight the trend of detectability as $CTDI_{vol}$ and MBIR level change. Points in the map with same colour shade are the dose-MBIR level combinations that have the same detectability.

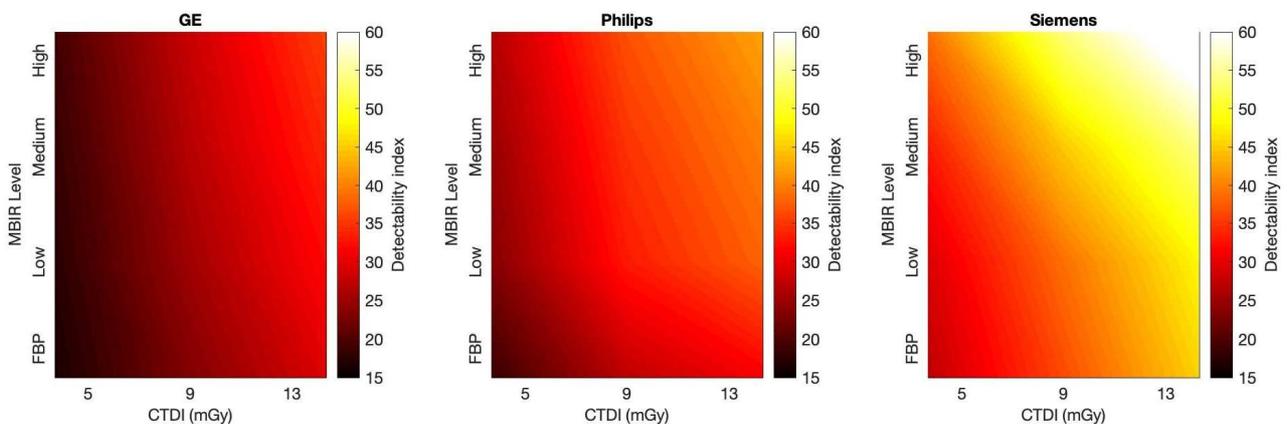

Fig. 5 Detectability index (d') as a function of the MBIR level (FBP, Low, Medium, High) and radiation dose level for a simulated lesion (size of 5 mm and |ΔHU|≈100 HU) for GE, Philips and Siemens scanner.

**Discussion**

The effectiveness of MBIR for dose reduction in CT examinations was demonstrated in several studies [34-39]. In this work, the evaluation was performed using state-of-the-art image quality metric that accounts for spatial resolution and noise reduction, together to clinical task and human observer characteristics. Setting side by side the performance of the three scanners, we found a great variability in image quality results. The main difference was in spatial resolution. GE and Siemens had similar $TTF_{50\%}$ values for the FBP reconstruction, while Philips had slightly lower values. The introduction of the partial-MBIR (GE) did not improve the FBP results.

Full-MBIR (Philips) increased the spatial resolution, without difference related to the iterative level, reaching the GE values. Different level of advanced-MBIR (Siemens) increased resolution gradually, achieving the highest $TTF_{50\%}$ values with the highest MBIR level. For each MBIR there was a gradual decrease of NPS area in conjunction with a shift of nNPS curve, which is related to the image texture, towards low spatial frequencies. This was more pronounced for full-MBIR followed by partial-MBIR and last advanced-MBIR. Siemens presented the best compromise since it had the lowest noise magnitude for each reconstruction and a limited noise spectrum impact. This was due to how the algorithm separates the noise from the anatomical structures [5] leading to a better appreciation by clinicians, because resulted in CT images with a less artificial "plastic" appearance [40].

Detectability plays a key role in image quality evaluation [41,42] since improvement of objective metrics [33] does not necessarily imply better diagnostic accuracy [43]. Therefore, as a second step, we exploited detectability a task-based image quality metric. The introduction of advanced-MBIR led to the most pronounced increase in d'. In absolute terms, Siemens showed the highest values of detectability. GE scanner presented a limited improvement both increasing dose and iterative level. Even the lowest value of d' obtained with Siemens (d'=27.12, 4.47mGy, FBP) could not be reached on GE by changing the iterative level but it needed a doubling of the dose, setting the highest iterative level. On Philips scanner, the same detectability could be obtained setting the highest level of MBIR or increasing the dose. Through d' we were able to quantify the potential dose reduction, to objectively rating CT image quality and to identify iterative level-dose combinations to optimise a protocol, considering a clinical task, with same d' on different scanners. This task-based image quality assessment has already been used to compare the performance or the reconstruction algorithms and define CT protocols [8,9,27-29,31,40]. Greffier systematically used d' as a metric for comparing different iterative algorithms but with results not fully in agreement with our data. Greffier assigned the lowest performance to partial-MBIR, followed by advanced-MBIR and then full-MBIR [31], while we found advanced-MBIR to be the best performing. This can be justified by the different choice of the clinical task, a fundamental factor for a task-based metric. We also went further, using this task-based assessment to compare different MBIR levels, at the contrary of Greffier who limited his study only to the middle iterative level. While on GE this did not lead to major differences in the results, because the performances changing the iterative level were remarkably similar, on Siemens the use of MBIR levels resulted in considerable dose savings.

Dose reduction in protocol optimisation must always consider a minimum image quality that depends on the clinical purpose [40,44]. With the introduction of d' or other similar metrics, it is possible to objectively define a quality standard that should not be lowered.

This study has some limitations. First, acquisitions were reconstructed using only a single-kernel and three iterative levels: another combination may lead to different result. Second, we evaluated only one task functions, which was representative of a task performed in clinical practice but is not the only one that should be considered. Finally, we did not compare the performance of NPWE with the human one, because NPWE model has been thoroughly validated against human observer performance [45,46].

In conclusion, task-based image quality was assessed for different dose and three increasing MBIR levels. Through this image quality metric, which is related to the clinical question, we rated image quality, evaluated the possible dose reduction and demonstrate that if we want to obtain equal d' for a specific protocol in different scanners, we must consider a combination of iterative and dose level, accepting that it is not possible to acquire always at the same dose value.

**References**

[1] Beister M, Kolditz D, Kalender WA Iterative reconstruction methods in X-ray CT. Phys Med. 2012. doi:10.1016/j.ejmp.2012.01.003

[2] Willemink MJ, Noël PB The evolution of image reconstruction for CT-from filtered back projection to artificial intelligence. Eur Radiol. 2019. doi:10.1007/s00330-018-5810-7

[3] Fan J, Yue M, Melnyk R Benefits of ASiR-V reconstruction for reducing patient radiation dose and preserving diagnostic quality in CT exams. 2014. White paper, GE Healthcare.

[4] Morton T Philips Iterative Model Reconstruction (IMR): Basic IMR testing, considerations and image quality trends. 2015. White Paper, Philips CT Clinical Science

[5] Ramirez-Giraldo JC, Grant K, Raupach R Admire Advanced Modeled Iterative Reconstruction. 2018. White Paper, Siemens Healthineers

[6] Arndt C, Güttler F, Heinrich A, Bürckenmeyer F, Diamantis I, Teichgräber U Deep Learning CT Image Reconstruction in Clinical Practice. Rofo. 2021. doi:10.1055/a-1248-2556

[7] Racine D, Becce F, Viry A et al Task-based characterization of a deep learning image reconstruction and comparison with filtered back-projection and a partial model-based iterative reconstruction in abdominal CT: A phantom study. Phys Med. 2020. doi:10.1016/j.ejmp.2020.06.004

[8] Greffier J, Frandon J, Si-Mohamed et al Comparison of two deep learning image reconstruction algorithms in chest CT images: A task-based image quality assessment on phantom data. Diagn Interv Imaging. 2021. doi:10.1016/j.diii.2021.08.001

[9] Greffier J, Dabli D, Frandon J et al Comparison of two versions of a deep learning image reconstruction algorithm on CT image quality and dose reduction: A phantom study. Med Phys. 2021. doi:10.1002/mp.15180

[10] Harding K, Thomson WH. Radiological protection and safety in medicine - ICRP 73. Eur J Nucl Med. 1997. 24(10):1207-9

[11] Vañó E, Miller DL, Martin CJ, et al ICRP Publication 135: Diagnostic Reference Levels in Medical Imaging. Ann ICRP 46. 2017. doi:10.1177/0146645317717209

[12] European Commission, Directorate-General for Energy, Jaschke W European study on clinical diagnostic reference levels for X-ray medical imaging: EUCLID. 2021. https://data.europa.eu/doi/10.2833/452154

[13] Siewerdsen JH, Cunningham IA, Jaffray DA A framework for noise-power spectrum analysis of multidimensional images. Med Phys. 2002. doi:10.1118/1.1513158

[14] Boedeker KL, Cooper VN, McNitt-Gray MF Application of the noise power spectrum in modern diagnostic MDCT: part I. Measurement of noise power spectra and noise equivalent quanta. Phys Med Biol. 2007. doi:10.1088/0031-9155/52/14/002

[15] Solomon JB, Christianson O, Samei E Quantitative comparison of noise texture across CT scanners from different manufacturers. Med Phys. 2012. doi:10.1118/1.4752209

[16] Li K, Garrett J, Ge Y, Chen GH Statistical model based iterative reconstruction (MBIR) in clinical CT systems. Part II. Experimental assessment of spatial resolution performance. Med Phys. 2014. 41(7): p. 071911. doi:10.1118/1.4884038

[17] Chen B, Christianson O, Wilson JM, Samei E Assessment of volumetric noise and resolution performance for linear and nonlinear CT reconstruction methods. Med Phys. 2014. doi:10.1118/1.4881519

[18] Solomon J, Samei E Quantum noise properties of CT images with anatomical textured backgrounds across reconstruction algorithms: FBP and SAFIRE. Med Phys. 2014. doi:10.1118/1.4893497

[19] De Marco P, Origgi D New adaptive statistical iterative reconstruction ASiR-V: assessment of noise performance in comparison to ASiR. J Appl Clin Med Phys. 2018. doi:10.1002/acm2.12253

[20] Maidment AD, Albert M Conditioning data for calculation of the modulation transfer function. Med Phys. 2003. doi:10.1118/1.1534111

[21] Richard S, Husarik DB, Yadava G, Murphy SN, Samei E Towards task-based assessment of CT performance: system and object MTF across different reconstruction algorithms. Med Phys. 2012. doi:10.1118/1.4725171

[22] Solomon JB, Wilson J, Samei E Characteristic image quality of a third generation dual-source MDCT scanner: noise, resolution, and detectability. Med Phys. 2015. doi:10.1118/1.4923172

[23] Yu L, Vrieze TJ, Leng S, Fletcher JG, McCollough CH Technical Note: Measuring contrast and noise dependent spatial resolution of an iterative reconstruction method in CT using ensemble averaging. Med Phys. 2015. doi:10.1118/1.4916802

[24] Vennart W. ICRU Report 54: medical imaging – the assessment of image quality. Radiography. 1996. 3:243–4.

[25] Burgess AE Statistically defined backgrounds: performance of a modified non-prewhitening observer model. J Opt Soc Am A Opt Image Sci Vis. 1994. doi:10.1364/josaa.11.001237

[26] Samei E, Bakalyar D, Boedeker KL et al Performance evaluation of computed tomography systems: Summary of AAPM Task Group 233. Med Phys. 2019. doi:10.1002/mp.13763

[27] Greffier J, Boccalini S, Beregi JP et al CT dose optimization for the detection of pulmonary arteriovenous malformation (PAVM): A phantom study. Diagn Interv Imaging. 2020. doi:10.1016/j.diii.2019.12.009

[28] Greffier J, Frandon J, Pereira F et al Optimization of radiation dose for CT detection of lytic and sclerotic bone lesions: a phantom study. Eur Radiol. 2020. doi:10.1007/s00330-019-06425-z

[29] Greffier J, Frandon J, Hamard A et al Impact of iterative reconstructions on image quality and detectability of focal liver lesions in low-energy monochromatic images. Phys Med. 2020. doi:10.1016/j.ejmp.2020.07.024

[30] The Phantom Laboratory. Catphan 500 and 600 Product Guide. 2021.

[31] Greffier J, Frandon J, Larbi A, Beregi JP, Pereira F CT iterative reconstruction algorithms: a task-based image quality assessment. Eur Radiol. 2019. doi:10.1007/s00330-019-06359-6

[32] Morisaka H Dose Reduction Strategies for Iodinated Contrast Agents: Low-Tube Voltage and Iterative Reconstruction. 1st edn, Springer, Cham. 2021. https://doi:10.1007/978-3-030-79256-5_5

[33] Verdun FR, Racine D, Ott JG, Tapiovaara MJ et al Image quality in CT: From physical measurements to model observers. Phys Med. 2015. doi:10.1016/j.ejmp.2015.08.007

[34] Katsura M, Matsuda I, Akahane M et al Model-based iterative reconstruction technique for radiation dose reduction in chest CT: comparison with the adaptive statistical iterative reconstruction technique. Eur Radiol. 2012. doi:10.1007/s00330-012-2452-z

[35] Yamada Y, Jinzaki M, Hosokawa T et al Dose reduction in chest CT: comparison of the adaptive iterative dose reduction 3D, adaptive iterative dose reduction, and filtered back projection reconstruction techniques. Eur J Radiol. 2012. doi:10.1016/j.ejrad.2012.07.013

[36] Yan C, Xu J, Liang C et al Radiation dose reduction by using CT with iterative model reconstruction in patients with pulmonary invasive fungal infection. Radiology. 2018. doi:10.1148/radiol.2018172107

[37] Laurent G, Villani N, Hossu G et al Teixeira PA Full model-based iterative reconstruction (MBIR) in abdominal CT increases objective image quality, but decreases subjective acceptance. Eur Radiol. 2019. doi:10.1007/s00330-018-5988-8

[38] Sun J, Yang L, Zhou Z et al Performance evaluation of two iterative reconstruction algorithms, MBIR and ASIR, in low radiation dose and low contrast dose abdominal CT in children. Radiol Med. 2020. doi:10.1007/s11547-020-01191-1

[39] Atri PK, Sodhi KS, Bhatia A, Saxena AK, Khandelwal N, Singhi P Model-based iterative reconstruction in paediatric head computed tomography: a pilot study on dose reduction in children. Pol J Radiol. 2021. doi:10.5114/pjr.2021.108884

[40] Racine D, Ryckx N, Ba A et al Task-based quantification of image quality using a model observer in abdominal CT: a multicentre study. Eur Radiol. 2018. doi:10.1007/s00330-018-5518-8

[41] Samei E, Richard S Assessment of the dose reduction potential of a model-based iterative reconstruction algorithm using a task-based performance metrology. Med Phys. 2015. doi:10.1118/1.4903899

[42] Solomon J, Zhang Y, Wilson J, Samei E An automated software tool for task-based image quality assessment and matching in clinical CT using the TG-233 Framework. Med Phys. 2018. 45/6:E134 - E134

[43] Solomon J, Marin D, Roy Choudhury K, Patel B, Samei E Effect of Radiation Dose Reduction and Reconstruction Algorithm on Image Noise, Contrast, Resolution, and Detectability of Subtle Hypoattenuating Liver Lesions at Multidetector CT: Filtered Back Projection versus a Commercial Model-based Iterative Reconstruction Algorithm. Radiology. 2017. doi:10.1148/radiol.2017161736

[44] Viry A, Aberle C, Lima T et al Assessment of task-based image quality for abdominal CT protocols linked with national diagnostic reference levels. Eur Radiol. 2021. doi:10.1007/s00330-021-08185-1.

[45] Solomon J, Mileto A, Ramirez-Giraldo JC, Samei E Diagnostic performance of an advanced modeled iterative reconstruction algorithm for low-contrast detectability with a third-generation dual-source multidetector CT Scanner: potential for radiation dose reduction in a multireader study. Radiology. 2015. doi:10.1148/radiol.15142005

[46] Racine D, Ba AH, Ott JG, Bochud FO, Verdun FR Objective assessment of low contrast detectability in computed tomography with Channelized Hotelling Observer. Phys Med. 2016. doi:10.1016/j.ejmp.2015.09.011